  \documentclass{gretsi}

\usepackage[english,french]{babel}   
  \usepackage{times}			
\usepackage{graphicx}
\usepackage{amsmath}
\usepackage{amsfonts}
\usepackage{algorithm}
\usepackage[noend]{algpseudocode}
 
\usepackage[T1]{fontenc}
\usepackage[utf8]{inputenc}

\DeclareMathOperator{\prox}{prox}

\def\B#1{\mathbf{#1}}

\title{FAASTA: A fast solver for total-variation regularization of
ill-conditioned problems with application to brain imaging}

\author{\coord{Ga\"el}{Varoquaux}{},
	\coord{Michael}{Eickenberg}{},
	\coord{Elvis}{Dohmatob}{},
	\coord{Bertand}{Thirion}{}}

\address{\affil{}{INRIA Saclay, \'Equipe Parietal \\
    Neurospin, B\^at 145, CEA Saclay, 91191 Gif sur Yvette, France}}

\email{firstname.lastname@inria.fr}

\frenchabstract{Il n'y a pas de formule analytique pour résoudre les
  problèmes de débruitage avec pénalité en variation totale, tout
  comme pour beaucoup d'autres problèmes de parcimonie en analyse. En
  conséquence, son utilisation pour régulariser un problème inverse
  conduit à de difficiles problèmes d'optimisation, qui sont souvent
  résolus par des méthodes de premier ordre. Cependant, lorsque le
  terme d'attache aux données est très mal conditionné et sans
  structure simple, comme en imagerie cérébrale, son optimisation est
  coûteuse. Il convient alors de minimiser le nombre d'itérations
  globales. Nous présentons pour cela fAASTA, une variante de FISTA
  qui utilise une optimisation interne pour l'opérateur proximal avec
  une tolérance adaptative. Nous illustrons son intérêt sur une étude
  empirique d'un problème de ``décodage cérébral''.  }

\englishabstract{The total variation (TV) penalty, as many other 
analysis-sparsity problems, does not lead to separable factors or a proximal operator
with a closed-form expression, such as soft thresholding for the $\ell_1$ penalty. 
As a result, in a variational formulation of an inverse problem or statistical
learning estimation, it leads to challenging non-smooth optimization problems
that are often solved with elaborate single-step first-order methods. When the
data-fit term arises from empirical measurements, as in brain imaging, it is
often very ill-conditioned and without simple structure. In this situation, 
in proximal splitting methods, the computation cost of the
gradient step can easily dominate each iteration. Thus it is beneficial
to minimize the number of gradient steps.
We present fAASTA, a variant of FISTA, that relies on an internal solver for
the TV proximal operator, and refines its tolerance to balance computational
cost of the gradient and the proximal steps. We give benchmarks and
illustrations on ``brain decoding'': recovering brain maps from noisy
measurements to predict observed behavior. The algorithm as well as the
empirical study of convergence speed are valuable for any non-exact proximal
operator, in particular analysis-sparsity problems.}

\begin{document}
\maketitle
\selectlanguage{english}

\section{Introduction: problem setting}

Minimizing a functional using the Total Variation (TV) of an image was
originally introduced for denoising purposes \cite{rudin1992nonlinear},
but it is useful in more general problems, as a regularization for
ill-posed inverse problems. For instance for image reconstruction in
computed tomography \cite{sidky2008image}, or in regression or
classification settings for brain imaging \cite{michel2011tv}. 
In image-recovery applications, unlike with pure prediction problems as in machine
 learning, optimizing the corresponding cost function
to a high tolerance is important (see \emph{eg} in brain imaging
\cite{dohmatob2014benchmarking}). Indeed, the penalty is most relevant in
the ill-conditioned region of the data-fit term. However, this
optimization is particularly challenging, as the TV penalty introduces a
long-distance coupling across the image in the final solution. This paper
studies optimization algorithms for TV-penalized inverse problems with an
ill-conditioned and computationally costly data-fit term, as in brain
imaging. In particular, we introduce a new variant of FISTA that is well
suited to inexact proximal operators.


The TV penalty can be seen as an instance of a wider set of
regularizations, analysis sparsity problems \cite{vaiter2014low}, that
impose sparsity on a linear transformation of the image, as \emph{eg} in
TV-$\ell_1$ used in brain imaging \cite{gramfort2013}.
In general, an analysis-sparsity risk-minimization problem is written as
\begin{equation}
    \hat{\B{w}} = \underset{\B{w}}{\text{argmin}}
    \biggl(\sum_{i=1}^n \mathcal{L}(\B{X}_i, \B{y}_{\!i}, \B{w})
    + \lambda \|\B{K}\, \B{w}\|_\bullet\biggr)
    \label{eq:analysis_sparsity}
\end{equation}
where $\mathcal{L}$ is the data-fit term, $\B{K}$ is the ``analysis
operator'', and
$\|{\cdot}\|_\bullet$ is a simple sparsity-inducing norm, such as the
$\ell_1$ or $\ell_{21}$ norm. Often, $\B{K}$ is rank-deficient, as in
over-complete dictionaries, and $\mathcal{L}$ is ill-conditioned. 
Problem (\ref{eq:analysis_sparsity}) is thus a challenging
ill-conditioned non-smooth optimization.

With linear models, $\B{X}$ is the design matrix. It is ill-conditioned
when different features are heavily correlated. These unfortunate
settings often happen when the design matrix results from experimental
data, as in brain imaging or genotyping, or for many measurement
operators that are nearly blind to some aspects of the weights $\B{w}$,
\emph{eg} the high frequencies in tomography or image deblurring
applications.
Analysis sparsity is then particularly interesting because it can impose
sparsity with $\B{K}$ specifically along those aspects of the weights.
For instance TV is edge-preserving: it sharpens some high-frequency
features.

\section{State of the art optimizers}



\sloppy
As the norm $\|.\|_\bullet$ is not smooth, standard gradient-based
optimization methods cannot be readily used to solve
(\ref{eq:analysis_sparsity}). Proximal-gradient methods \cite{parikh2013}
generalize the
gradient step with an implicit subgradient step using the
\textit{proximal operator}, defined by
\mbox{$\prox_{g}(\B{y}) = \text{argmin}_\B{x}\frac{1}{2}\|\B{y} - \B{x}\|_2^2 +
g(\B{x})$}.

\paragraph{Proximal iterations}
If $f$ is the data-fit term, which we assume smooth, while $g$ is the
non-smooth penalty, the simplest method is the 
Iterative Shrinkage-Thresholding Algorithm (\textbf{ISTA})
\cite{daubechies2004iterative,combettes2011proximalsplitting}: alternating gradient descent on
the smooth part and application of the proximal map on the non-smooth
part with iterations of
\begin{equation}
    \B{w}_{k + 1} = \prox_{\frac{1}{L} g}\smash{\left(\B{w}_k -
    \frac{1}{L}\nabla_{\!\B{w}} g(\B{w}_k)\right)},
    \label{eq:ista_iteration}
\end{equation}
where \(L\) is the Lipschitz constant of \(\nabla_{\!\B{w}} f\). 
An accelerated gradient method, that is multi-step, can speed up
convergence for ill-conditioned problems by adding a momentum term:
in the \textit{fast iterative shrinkage-thresholding algorithm}  
(\textbf{FISTA}) \cite{beck09afast} the gradient steps are 
applied to a 
combination of $\B{w}_k$ and $\B{w}_{k - 1}$.
The drawback is that there is no guarantee that each step of FISTA
decreases the energy and large rebounds are common. This
non-monotone behavior can be remedied by switching to ISTA
iterations whenever an increase in cost is detected, as in 
\textit{monotone FISTA} (\textbf{mFISTA})
\cite{beck09fastgradient-based}.

\paragraph{Proximal algorithms for analysis sparsity}
The success of ISTA-type algorithms hinges upon the computation of the
proximal operator, that is very efficient for synthesis sparsity as in
lasso and elastic-net like problems ($\ell_1$ and $\ell_1 + \ell_2$
penalties). However, for many analysis-sparsity proximal operators there is 
no analytical formula. If the norm used in the analysis sparsity is
$\ell_1$ or an $\ell_{21}$, writing the dual problem leads to a constraint
formulation amenable to an accelerated projected gradient
\cite{beck09fastgradient-based}, as the dual norms are respectively an
$\ell_\infty$ and $\ell_{2\infty}$ with easy projections to the unit
ball. Importantly, a monotone scheme can also be used, and the tolerance of
the optimization can be controlled via the dual gap \cite{michel2011tv}.

While the computation of the proximal operator is no longer exact,
proximal gradient algorithms can be shown to converge as long as error on
the proximal operation decreases sufficiently with the iteration number
$k$ of the outer loop \cite{schmidt2011convergence}. To achieve the best
convergence rates in $k$, the error may be required to decrease as fast
as $\varepsilon_k \sim k^{-4}$. Using an accelerated algorithm to compute
the proximal, achieving a tolerance of $\varepsilon$ is in
$\mathcal{O}(\sqrt{\varepsilon})$. Thus, in such a scheme the cost of
each iteration increases as $k^{2}$, which renders optimization to a
tight tolerance prohibitive.

\paragraph{Operator splitting algorithms}
The costly inner loop is required in ISTA-type algorithms because the
proximal operator cannot be easily computed on the primal variables of
the optimization. Another family of proximal algorithms rely on
``splitting'': introducing auxiliary variables in the analysis space:
$\B{z} = \B{K}\, \B{w}$. Optimizing on these is then a simple $\ell_1$ or
$\ell_{21}$ proximal problem. The alternating direction method of
multipliers (\textbf{ADMM}) is such an approach that is popular for
signal processing applications for its versatility \cite{boyd2011}. Note
that it relies on the choice of a hyper-parameter, $\B{\rho}$ controlling
the tightness of the split that can have an impact on the convergence.
Similarly, the primal-dual algorithm in \cite{chambolle-pock:2011} relies on two
hyperparameters and can be written as a preconditioned version of ADMM.


\paragraph{Optimizing a smooth surrogate}
Newton or quasi-newton algorithms are excellent optimizers for
smooth problems that achieve quadratic convergence rates under weak
conditions and are somewhat unaffected by ill conditioning
\cite{nocedal2006numerical}. In particular, the limited-memory BFGS
(\textbf{LBFGS}) \cite{liu1989limited} quasi-newton scheme is well suited
to high-dimensional problems.
An approach to minimizing a convex non-smooth function relies on
minimizing series of smooth upper bounds, decreasing the approximation as
the algorithm progresses
\cite{dohmatob2014benchmarking,dubois2014predictive}.
Intuitively, the benefit of such an approach is that it should be
well suited to strong ill-conditioning of the smooth part of the energy,
to the cost of more difficulty in optimizing the non-smooth part.

\paragraph{Iteration cost}
When the design matrix $\B{X}$ is large and dense, as in brain imaging or
genomics, the computation cost of multiplying it with a vector or matrix
is by far the most expensive elementary operations as  does not fit in
$\B{X}$
CPU cache. Thus it is important to limit as much as possible these
operations, which typically come into play when computing the gradient or
energy of the smooth data-fit part of the problem\footnote{Note that such
situation is different than in many signal-processing applications in
which $\B{X}$ is a sparse or structured operator as a convolution or a
wavelet transform, and thus can be applied very fast.}.

\section{FAASTA: adapting inner tolerance}

ISTA-type schemes are parameter-free algorithms with a good convergence
rate. However their optimal use in analysis-sparsity problems require
balancing the computational cost of the gradient step with that of the
proximal operator. Indeed, while a tight tolerance on the proximal
operator may guarantee the optimal convergence speed in terms of number
of iterations of the outer loop, it implies increasing computing time for
each iteration. Thus, for a low desired tolerance on the global energy
function, it may be beneficial to set a lax tolerance on the dual gap of
the proximal operator, and thus achieve fast iterates. On the opposite,
optimizing the global problem to a stringent tolerance requires very high
precision on the proximal operator, which in turn slows down iterations
of the outer loop. 

\algrenewcommand\algorithmicindent{.8em}
\begin{algorithm}[t]
    \caption{fAASTA}\label{algo:faasta}
    \begin{algorithmic}
    \Require \(\B{w}_0\)
    \State \(\text{ISTA}\gets False,\qquad\)
    \(\B{v}_1\gets \B{w}_0,\qquad\)
    \(k\gets 0,\qquad\)
    \State \(t_1 \gets 1,\qquad\)
    \(dgtol\gets 0.1\)\;
    \While{not converged}
	    \State \(k \gets k + 1,\quad\)
	    \State \(\B{w}_{k} \gets \prox_{\frac{g}{L}} (\B{v}_k - \frac{1}{L}\nabla
		f(\B{v}_k), dgtol)\)\;
	    \If{\(\mathcal E(\B{w}_k) > \mathcal E(\B{w}_{k - 1})\)}
		    \State \(\B{w}_k\gets \B{w}_{k - 1},\qquad\)
		    \(\B{v}_k\gets \B{w}_{k - 1}\)\;
		    \If{ISTA}
			\State \(dgtol\gets dgtol / 5\)\;
			\While{
			\State\(\mathcal E(
			\prox_{\frac{g}{L}} (\B{v}_k -
\frac{1}{L}\nabla_{\!\B{w}} f(\B{v}_k),
			dgtol)) > \mathcal E(\B{w}_{k-1})\)}
				\State \(dgtol\gets dgtol / 5\)
			\EndWhile
		    \EndIf
		    \(\text{ISTA}\gets True\)\;
	    \Else
		    \If{ISTA}
			\State \(\B{v}_k\gets \B{w}_k, \qquad\)
			 \(\text{ISTA}\gets False\)
		    \Else\vspace*{-.5ex}
			\State 
			    \(t_k\gets\frac{1 +
				\sqrt{1 + 4t_{k-1}^2}}{2},\qquad\)
			 \State \(\B{v}_k\gets \B{w}_k + 
				\frac{t_{k-1} - 1}{t_k}
				    (\B{w}_k - \B{w}_{k - 1})\)\;
		    \EndIf
		    
	    \EndIf
    \EndWhile
    \end{algorithmic}
\end{algorithm}

We introduce a new variant of ISTA, fast Adaptively Accurate Shrinkage
Thresholding Algorithm (\textbf{fAASTA}). The core idea is to adapt the
tolerance of the proximal operator as needed for convergence. For this,
we rely on the fact that in an ISTA iteration
--eq.\,(\ref{eq:ista_iteration})-- the global energy must decrease
\cite{daubechies2004iterative}. Thus we adapt the tolerance of the
proximal-operator solver to ensure this decrease. In practice, in the inner loop,
we control that tolerance by checking the dual gap of the proximal
optimization problem, which is much cheaper to compute than the global
energy, as it does not involve the accessing the large, dense design
matrix $\B{X}$. We decrease this dual gap setting by a factor of 5 when
we observe that an ISTA iteration did not result in an energy decrease.
In addition, to benefit from accelerated schemes as in FISTA, we rely on
an mFISTA strategy: we apply FISTA iterations but when the energy
increases, switch to an ISTA step, in which we can then, if needed,
decrease the tolerance on the inner dual gap.

The actual procedure is described in alg.\,\ref{algo:faasta}. One
difficulty is implementing the back and forth between ISTA and FISTA
without recomputing intermediate variables. With speed in mind, another
important implementation aspect is to factor out expressions common to
the gradient and energy computations in order to access the large $\B{X}$
matrix as little as possible. These breakdowns are not detailed here.

\section{Empirical study: TV-$\ell_1$ for decoding}

Here, we benchmark the various algorithms on a brain-decoding
application: the task is to predict subject's behavior, encoded as a
categorical variable in $\B{y}$, from fMRI images that constitute the
design matrix $\B{X}$. For this purpose, we use a TV-$\ell_1$-penalized
logistic regression, which is written as an analysis sparsity problem
in notations of eq.\,(\ref{eq:analysis_sparsity}):
\begin{align}
\rlap{\hspace*{-1.8ex}\text{loss}} \;\;& & 
    \mathcal{L}(\B{X}_i, \B{y}_{\!i}, \B{w}) &=
\log\bigl(1 + \exp\bigl(-\B{y}_{\!i}(\langle \B{X}_i, \B{w}\rangle + b)\bigr)
\bigr)\hspace*{-1ex}&
\\
\rlap{\hspace*{-1.8ex}\text{penalty}} \;\;& & \|\B{K}\, \B{w}\|_\bullet &=
(1 - \alpha)\| \nabla \B{w}\|_{21} + \alpha\|\B{w}\|_1\hspace*{-1ex}&
\end{align}
where $\nabla$ is the image-domain finite difference operator and the
$\ell_{21}$
norm groups the different image directions for an isotrope TV
formulation. 
We set $\alpha = .25$.
We investigate discrimination of whether the subject is presented
images of shoes or scrambled pictures in a study of human vision, using
the data from \cite{haxby2001}.
We rely on the nilearn library for data download, loading and cleaning
\cite{abraham2014machine}.


We study convergence of the following approaches: ISTA and mFISTA schemes
with \emph{i)} a lax dual-gap tolerance ($.1$) on the inner problem,
\emph{ii)} a tight dual-gap tolerance ($10^{-10}$), \emph{iii)} an
adaptive control of the dual-gap tolerance as exposed above, and
\emph{iv)} the
decrease motivated theoretically in \cite{schmidt2011convergence};
an LBFGS optimizer on a series of smooth upper-bounds following the
approach of \cite{dohmatob2014benchmarking}; an ADMM with different
values of the $\rho$ control parameter (implemented as in 
\cite{dohmatob2014benchmarking}).

\begin{figure*}[bt]
\includegraphics[width=.365\linewidth]{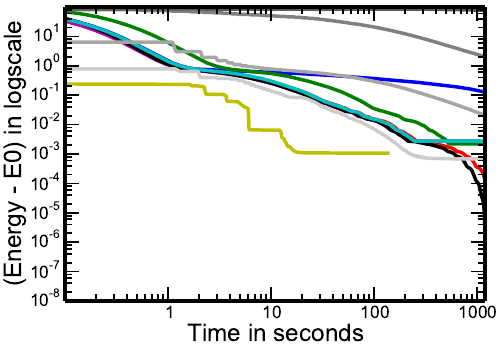}%
\llap{\raisebox{.14\linewidth}{%
    \rlap{$\lambda = 0.0143$}\hspace*{.3\linewidth}}}%
\hfill%
\hspace*{-.05\linewidth}%
\includegraphics[width=.365\linewidth]{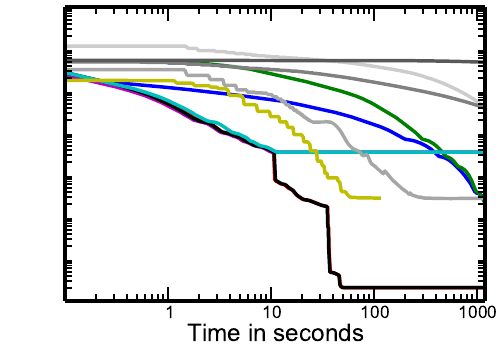}%
\llap{\raisebox{.14\linewidth}{%
    \rlap{$\lambda = 1.43$}\hspace*{.3\linewidth}}}%
\hfill%
\hspace*{-.05\linewidth}%
\includegraphics[width=.365\linewidth]{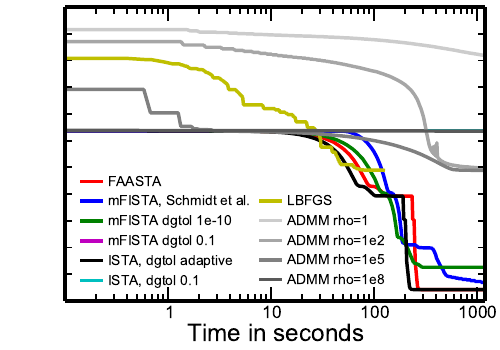}%
\hfill%
\llap{\raisebox{.135\linewidth}{%
    \rlap{$\lambda = 14.3$}\hspace*{.3\linewidth}}}%
\caption{\textbf{Convergence of the different optimization algorithms},
for 3 scenarios, with weak, medium and strong regularization, where
medium regularization corresponds to the value chosen by
cross-validation. These are log-log plots with the 0 defined as the
lowest energy value reached across all algorithms.
\label{fig:convergence}%
}
\end{figure*}

For all algorithms, we investigate convergence for 3 different values
of the regularization parameter, centered around the value minimizing
prediction error as measured by cross-validation.
On fig.\,\ref{fig:convergence}, we report energy as a function of time.
Importantly, we report time, and not iteration number. Indeed, not only
is computing time the most important quantity for the user, but also it
is suitable to compare different algorithms.

\bigskip

We find that, as expected, ISTA or FISTA with a lax tolerance on the
inner problem reaches an energy plateau before convergence although, for
weak to medium regularization, it progresses fast before hitting this
limitation. On the opposite, setting a stringent tolerance leads to a
slow but steady decrease of the energy (though it eventually also
reaches a plateau before complete convergence in the high-regularization
case). The theoretical decrease on the tolerance has an intermediate
behavior. In particular, it does not stagnate. The convergence of the
ADMM approach depends strongly on its control parameter: on the
one hand, no control parameter achieves satisfying convergence for all
regularization values, on the other hand the ADMM does not converge at
all for a bad choice of the parameter. LBFGS reaches a plateau, but gets
there fast in the case of weak regularization. Finally, our adaptive
approach, whether in its accelerate form, or simply in an ISTA loop,
reaches the lowest energy value.

\section{Discussion and conclusions}

Optimizing an analysis-sparsity functional with a ill-conditioned
data-fit term is a very challenging problem as the non-smooth part cannot
be readily written as a proximal problem with an analytic solution.
Splitting approaches with auxiliary variables in the analysis space are
often put forward as a solution for these problems as the non-smooth
contribution is easy to solve on these variables. However, when the
computation cost of the smooth data-fit part is high, \emph{eg} when the
design matrix is dense and large as in brain imaging, our experience is
that these approaches are sub-optimal --see also
\cite{dohmatob2014benchmarking}, that explored the preconditioned variant
of \cite{chambolle-pock:2011}. We conjecture that in these situations,
the choice of the control parameter becomes crucial to accelerate the
convergence\footnote{We evaluated the heuristics describe in
\cite{boyd2011} to evolve the control parameter, but with no conclusive
success.}.

Our preferred solution relies on using an inexact proximal operator and
adapting ISTA-type algorithms to set the corresponding tolerance.
Experiments show that this approach is robust and always converges to a
high tolerance. For very low regularization, the smooth data-fit part of
the functional dominates and careful use of smooth optimization methods
can achieve a quicker initial rate of convergence but are eventually
limited. We currently lack theory on the optimal strategy to set the
tolerance of the inner problem. Indeed, on the experiment studied in this
contribution, the accelerated gradient variant of the algorithm did not
outperform the single-step variant.

Analysis sparsity can be used to capture structure in a much wider set of
situations than the simpler, more common synthesis sparsity. However it
is seldom used in practice, likely because of the computational cost it
entails. Many problems can benefit from faster solvers outside of brain
imaging, for instance in tomography image recovery \cite{sidky2008image}.

%

\bibliographystyle{abbrv} 
\bibliography{biblio}

\end{document}